\documentclass[epsfig,12pt]{article}
\usepackage{epsfig}
\usepackage{amsfonts}
\begin {document}

\title {LINEARIZED MULTIPOLE SOLUTIONS AND THEIR REPRESENTATION}

\author{J.L. Hern\'andez-Pastora\thanks{ETS Ingenier\'\i a Industrial de
B\'ejar. e-mail address: jlhp@usal.es} \\
Departamento de Matem\'atica Aplicada and IUFFyM\thanks{Instituto Universitario
de F\'\i sica Fundamental y Matem\'aticas.} \\
Universidad de Salamanca.  Salamanca,
Spain  \\}

\date{\today}

\maketitle

\vspace{-1cm}

\begin{abstract}
The monopole solution of the Einstein vacuum field equations
(Schwarzschild`s solution)  in Weyl coordinates involves a metric function that
can be interpreted as the
gravitational potential of a bar of length $2m$ with constant linear density.
The question addressed in this work is whether similar representations can be
constructed for Weyl solutions other than
the spherically symmetric one.

A new family of static solutions of the axisymmetric vacuum field equations
generalizing the M-Q$^{(1)}$ solution \cite{mq1} is developed.
These represent slight deviations from spherical symmetry in terms of the
relativistic multipole moments (RMM) we wish
 the solution to contain. A Newtonian object referred to as a dumbbell can be used to
describe these solutions in a simple form by means
of the density of this object, since the physical properties of the relativistic
solution are characterized by its behaviour.
  The density profile of the dumbbell, which is given in terms of the RMM of the
solution,
allows us to distinguish general multipole Weyl solutions from the
constant-density Schwarzschild solution.
The  range of values  of the multipole moments that generate  positive-definite
density profiles are also calculated.
The  bounds on the multipole moments that arise from this density condition are
identical to those required for a well-behaved
 infinite-redshift surface $g_{00}=0$.
\end{abstract}


\newpage

\section{Introduction}
As  is well known, all the static solutions, with good asymptotical behaviour,
of the axially symmetric  Einstein vacuum equations
 are given by the Weyl family of solutions \cite{weyl}. Each solution of this
family is characterized by means of a particular
set of coefficients $\{a_n\}$ through a series that provides the metric function
of the solution in Weyl coordinates $\{r,\theta\}$ ($P_l$ being the  Legendre
polynomials)
\begin{equation}
  \Psi=\sum_{n=0}^\infty\frac{a_n}{r^{n+1}}P_n(\cos\theta) \ .
\label{psi}
\end{equation}
Several works  have been devoted \cite{bakdal}, \cite{mq1}, \cite{11},
\cite{mq} to establishing a relationship between
this set of Weyl coefficients and the Relativistic Multipole Moments (RMM)
of the solution, in an attempt to provide a mechanism for selecting solutions
with physical meaning from the Weyl family. Naturally, the Schwarzschild
 solution, which is  spherically symmetric, belongs to that
family and its set of Weyl coefficients (\ref{aeschw}) is known \cite{tesis}.
Alternatively, the Erez-Rosen-Quevedo (ERQ) representation
\cite{erq}, \cite{forstchrite} of the static and axisymmetric  vacuum solutions in prolate
spheroidal coordinates
has another set of coefficients $\{q_n\}$ (obviously related with $\{a_n\}$; see
 \cite{tesis} for details), whose values associated with the Schwarzschild solution
are
$\{q_0=1, q_n=0 \  \forall n \}$. An evident advantage of this representation is
a simpler form of the metric function that describes
 the spherically symmetric solution. In \cite{msa} the
suitability
of these coordinates was discussed for describing  spherical symmetry in General Relativity
(GR)
because the coordinates are harmonic for that class of metrics.

The identification of a given solution among all the solutions from either
the Weyl family or  the ERQ representation is a procedure that is
neither unique nor easy to develop, despite the work done. That is due to the
difficulties involved in establishing  a relationship between
 the RMM of the solution and the Weyl
coefficients, specially 
for solutions with a relevant physical meaning. In addition, the
interpretation of a solution, once  constructed or chosen
from the Weyl family,
 is not  provided by a well-known and definitively established procedure, unless
the solution is constructed with  the desired RMM.
In this case, the parameters of a particular solution
acquire a physical meaning by means of the
RMM.
Nevertheless, we have to deal with a set of infinite coefficients and quite complicated
expressions for the metric functions of the line element.

The aim of the paper is dual: first, we develop a class of
solutions of the Weyl
family
with a very interesting physical meaning -the so called {\it Linearized
Multipole} (LM)
Solutions-  because they are
constructed as the solutions that have a fixed and finite number ($g+1$) of
RMM. The assumption that all those multipoles are  quantities of a
very small magnitude is taken into account. With this approach, used for   the calculation of the
specific
Weyl coefficients, we obtain  exact solutions of the static and axisymmetric
 Einstein vacuum equations with the physical meaning explained above. The M-Q$^{(1)}$ solution \cite{mq1} 
 is recovered for the case $g=1$. We gain  a new solution
 that is able to describe relativistic gravitational fields very close to the
Schwarzschild solution and hence it might become a very useful tool for researching
the eventual physical effects due to deviations from spherical
symmetry \cite{luisgyr}, \cite{luisgeod}.ç

Second, we shall attempt to give an interpretation of these solutions by means of
the gravitational potential of a Newtonian object.
As is known, the Schwarzschild
solution can be viewed as a bar of length $2m$ with constant linear density
since the gravitational potential of such an object equals the metric function
of the Monopole solution. Although this representation of a spherically-symmetric GR solution
is disconcerting, it at least provides us with a characterization
of the Monopole solution by means of the gravitational field of a body that in Newtonian Gravity (NG)
would be a spherical distribution of mass $m$. The
role
played by that  spherical distribution in NG is  adopted by a non-spherical object (the
constant-density bar) to describe  the space-time with spherical symmetry in GR.
The questions addressed here are as follows: How could  a  solution of the Weyl family be identified
with some  gravitational potential?  Is there any physical object available to depict the solution defined
by means of the  metric function $\Psi$? What  characteristics of that object are
appropriate for  representing  that solution and what can they be used for?
If  we identify the spherically symmetric solution with a bar of constant linear density, what could we identify the other solutions with?

In \cite{letelier} Letelier  gives a physical interpretation of the multipolar
ERQ solutions of the Einstein vacuum equations in terms of bars, and the author
finds that each {\it "multi-pole"} corresponds to the Newtonian potential  of a
bar  with a linear density proportional to a Legendre polynomial. First of all,
we
must  point out that what he calls {\it "multi-pole"} has nothing to do with a
solution of the Einstein equations  possessing  a single RMM, except for the
Monopole case. However, this work establishes a relationship between each
coefficient $\{q_n\}$ of the ERQ representation and a bar with a certain linear
density. Since each static and axisymmetric vacuum solution can be written in
the ERQ representation, we can obtain an interpretation of the series
that defines the solution as the infinite sum of potentials associated with bars
of equal length and certain densities \cite{letelier}.
Nevertheless, we shall see that this definition has important limitations and it
 proves to be
useless for constructing  a linear density of a single bar that allows us to
describe the whole series of the solution in general cases.

The question of looking for an interpretation of the metric function $\Psi$
is developed in the case of the LM solutions. The conclusion we
obtain leads to the adoption of an object  slightly different to a single bar that
plays the role of being the Newtonian representation
of  these solutions: A {\it dumbbell}, i.e., an
object consisting of a
bar and two identical masses, one at each end of the bar, will identify the LM
solutions. The gravitational potential of such an object provides the metric function of the LM solutions 
by conveniently changing the linear density of the bar as well as the
magnitudes of the masses. It must be pointed out here
that
this work provides a method for calculating both the linear density and the
masses of the dumbbell in terms of the RMM possessed by the LM solution. Therefore, the differences with respect to the Schwarzschild
solution, which is recovered from the LM solution by taking all RMM higher than
the
monopole equal to zero,  can be specified explicitly. A detailed study of the
bounds on the magnitude of the RMM imposed by the condition of a well defined positive
linear density is carried out.

Finally we wish to stress some reasons why we think that this representation is
useful and physically interesting.
The relevant metric function of the LM solution is calculated from the density
of the dumbbell, which is an even polynomial of $g$ degree. Therefore,
 the complete solution is defined by  $g$ independent coefficients alone. This is a
very simple way of describing a family of static and
axisymmetric vacuum solutions rather than
the Weyl or the ERQ
representations. With the dumbbell we are introducing a generalization of the
Newtonian representation of the Schwarzschild solution.
Furthermore, we can use the dumbbell to describe some physical properties of
the relativistic solution. We shall show that the
 positive-definite condition of the density  allows us to decide whether  the
source that generates the space-time can be described as a
 flattened or elongated isolated body
with respect to the spherical mass distribution. In addition, the existence of a
connected horizon on the
 infinite-redshift surface $g_{00}=0$  can be described by means of a
well-behaved density condition.

The paper is organized as follows: in section 2  we explicitly calculate the
expression for the
Newtonian
multipole moments of  a bar of finite length with a
linear density as well as the gravitational potential of such a mass
distribution.  In subsection 2.2 we first analyze the
definition introduced by Letelier \cite{letelier} and we claim the existence
of a Newtonian representation
 of the Weyl solutions in analogy with the case of  spherical symmetry.

In section 3 we present the LM solution. Its metric function and its Weyl
coefficients are shown, and we introduce the dumbbell as a suitable object  to
represent this solution. We explicitly calculate its characteristics (density
and masses) in terms of the RMM. The section is completed with a detailed
analysis of the properties of the dumbbell and the calculation of its
gravitational potential. Section 4 is devoted to showing some examples of special
physical interest: we make a comparison between their linear densities and we
look for bounds on the values of the RMM. The Erez-Rosen solution, the M-Q$^{(1)}$
solution and a particular case of the LM solution are studied. A Conclusions section
reports an analysis of the results obtained and discussions about their relevance, together with future
 extensions of the work.
We also include an Appendix that is used to show  that the definition of density
introduced by Letelier can be recovered by an
alternative procedure.

\section{Description and interpretation of relativistic solutions by means of a
Newtonian gravitational potential and its multipole moments}

\subsection{Newtonian gravity}

Let us consider the Newtonian gravitational potential of a mass distribution
with density
$\rho(\vec {\hat z})$, given by the following solution of the Poisson equation (we use units in which the gravitational constant $G=1$)
\begin{equation}
\Phi (\vec x) = - \int_V\frac 1R \mu(\vec {\hat z}) d^3\vec {\hat z} \ ,
\label{uno}
\end{equation}
where the integral
is extended over the volume
of the source, $\vec {\hat z}$ is the vector that gives the position of a
generic point inside the source,
and
$R$ is the distance between that point and any exterior point
$P$, which is defined by its position vector $\vec x$. Let us now make an expansion of
this potential in  a power series of the inverse of the distance from the origin
to the point $\vec P$ ($r \equiv{\cal j}\vec x {\cal j}$) by means of a Taylor
expansion of the term
$\displaystyle{\frac{1}{R}}$ around the origin of coordinates, where $R\equiv
\sqrt {(x^i-{\hat z}^i)(x_i-{\hat z}_i)}$. For the case of an axially symmetric
mass distribution, this multipole development leads to a
Newtonian potential with the same form as equation (\ref{psi}) but with the Weyl
coefficients $\{a_n\}$ replaced by $-M_n^{NG}$, which
are parameters that denote  the massive multipole moment of order $n$ that can
be
defined by means of an integral expression extended
to the volume of the source,
\begin{equation}
M_n^{NG} = 2 \pi \int\int {\hat z}^{n+2}\mu({\hat
r},\hat\theta)P_n(\cos\hat\theta)
\sin\hat\theta d\hat\theta d{\hat r} \ ,
\label{moment}
\end{equation}
${\hat r}\equiv |\vec{\hat z}|$ representing the radius of the integration point
and
$\hat\theta$ the corresponding polar angle.

From the next section onwards we shall make use of a source  such as a bar of
length $2L$, centered and located along
the $Z$ axis. There is a well-established framework in Newtonian Gravity (NG) for
handling distributional line-sources such as this,
and we shall therefore consider
 an object described by a line singularity on the $Z$ axis with the following
linear density:
\begin{equation}
 \mu(\vec {\hat z})= \frac{1}{2\pi}\frac{\delta(\hat \rho)}{\hat \rho}\mu({\hat z}) \ ,
\label{mudelta}
\end{equation}
for some non-negative function $\mu(\hat z)$, $\delta(\hat \rho)$ being the Dirac
function  $\delta(\hat \rho-\hat \rho_0)$ at $\hat \rho_0=0$ and where  $\{\vec {\hat z}\}
\equiv \{{\hat \rho},{\hat z}\}$ are cylindrical coordinates.
Consequently, from equation (\ref{uno}) the gravitational potential of
such a  mass distribution is as
follows:
\begin{equation}
\Phi (\vec x) = - \int_V\frac 1R \mu(\vec {\hat z}) d^3\vec {\hat z} =
-\int_{-L}^L
\frac{\mu({\hat z})}{\sqrt{\rho^2+(z-{\hat z})^2}} d{\hat z}\ ,
\label{phiNG}
\end{equation}
where the position vector $\vec x$ is given by coordinates $(\rho,z)$, and $\vec
{\hat z}$ is located along the $Z$ axis.

According to equation (\ref{moment})  the Newtonian multipole
moments of this object (if the function $\mu(\hat z)$ is even in $\hat z$) are
as follows\footnote{An identical conclusion can be derived if we approximate the source by a series of
cylinders of radius $\epsilon$ and then take the limit
$\epsilon \rightarrow 0$ or, equivalently, we consider the factor $\epsilon/L$
negligible since we are dealing with a very narrow bar  ($\epsilon << 2L$).}:
\begin{equation}
 M_{2n}^{NG}=\int_{-L}^L  {\hat z}^{2n} \mu({\hat z}) d{\hat
z}=L^{2n+1}\int_{-1}^1 X^{2n} \mu(LX)dX \ .
\label{mNG}
\end{equation}

\subsection{Interpretation of  axisymmetric  and static solutions of
the   Einstein vacuum equations}

As is known, the line element of  a static and axisymmetric   vacuum space-time
is represented in Weyl form as follows
\begin{equation}
ds^2 = -e^{2\Psi} dt^2 +e^{-2\Psi}\left[
e^{2\gamma}\left(d\rho^2+dz^2\right)+\rho^2 d\varphi^2\right]
\end{equation}
where $\Psi$ and $\gamma$ are functions of the cylindrical coordinates $\rho$ and
$z$ alone. The metric function $\Psi$ is a solution of the Laplace equation
($\triangle \Psi=0$), and the other metric function $\gamma$ satisfies a system
of differential equations whose integrability condition is merely the equation for
the function $\Psi$. The Weyl family of solutions with good asymptotic
behaviour are given in spherical coordinates $\{r, \theta\}$ as the series
(\ref{psi}).

From the point of view of understanding the relativistic properties of the line
element corresponding to any particular choice of the potential function $\Psi$,
what is important,  of course, is not the Newtonian moments $a_n$ but the
RMM of the solution, first defined for static
and axisymmetric vacuum solutions by Geroch \cite{geroch} and Thorne
\cite{thorne}. The Newtonian moments were first expressed as functions of the
RMM by Fodor, Hoenselaers and Perj\'es \cite{fhp}, and by B\"ackdahl and
Herberthson later \cite{bakdal},  \cite{sueco}. Although the full FHP relations
are extremely complicated, they can be used to obtain relatively simple formulas
for the coefficients $\{a_n\}$ in situations where the deviation of the
relativistic solution from spherical symmetry is small. Some authors have devoted
their work to this research \cite{tesis}, \cite{mq1}, \cite{sueco}. In this sense,
the spherically symmetric line element (i.e. the Schwarzschild solution) only
has one RMM, the Monopole, and its metric function $\Psi$ is represented in
Weyl coordinates by a bar
of constant linear density of length $2M$, $M$ being the mass of the Monopole
solution. This interpretation is derived
from the fact that the metric function $\Psi$ of the solution is equal to the
Newtonian
gravitational potential of such an
object, i.e, if we take $\mu({\hat z})=1/2$ in (\ref{phiNG})
we obtain the metric function of the Schwarzschild solution in Weyl
coordinates \cite{tesis}
\begin{equation}
 \Psi=\Phi=\frac 12 \ln
\left(\frac{z+M-\sqrt{\rho^2+(z+M)^2}}{z-M-\sqrt{\rho^2+(z-M)^2}}\right) \ .
\label{both}
\end{equation}

Moreover, it is also known \cite{tesis}, \cite{mq} that the set of infinite
coefficients $\{a_n\}$
corresponding to the Schwarzschild solution
are\footnote{Note that the expression for these coefficients can easily be verified
by developing the metric function (\ref{both})
in power series of $1/r$ and identifying the expression with (\ref{uno}).}
\begin{equation}
 a_{2n}=-M^{2n+1}/(2n+1) \ , \ a_{2n+1}=0 \ .
\label{aeschw}
\end{equation}
These quantities are
precisely equal (up to a sign) to the Newtonian multipole moments $M_{2n}^{NG}$ of
the bar with
linear density $\mu=1/2$, and they can be explicitly
calculated\footnote{Let us note that we have to take the half-length of the
bar $L$ equal to $M$.} from
expression (\ref{mNG}). Therefore, the identification between the metric
function of the relativistic solution and the
gravitational potential of the Newtonian mass distribution leads to a curious
interpretation of the spherically symmetric solution of the
 Einstein vacuum equations. Let us remember that if $\Psi$ is generated by a
Newtonian point particle with mass M lying at the origin of coordinates so that
it has the spherical form $\Psi=-M/\sqrt{\rho^2+z^2}$, then the corresponding
line element  describes the Curzon metric \cite{curzon}, which is not
spherically symmetric.

In spite of the rather strange nature of this representation of
the
Monopole
Solution in GR,
very distant from Newtonian common sense,
this identification with a Newtonian
object  may be useful to characterize solutions of the Weyl family, and at the same time
it reveals how different the
multipole moments
are from NG to GR. Hence, we can assume that a physical object whose Newtonian
moments equal the  coefficients $\{a_n\}$ of the Weyl solution  allows us to
identify  this solution since the Newtonian gravitational potential $\Phi$ of
the object resembles  the metric function $\Psi$ of the relativistic solution.
The question
arising here is whether we could arrange a similar
identification for any other Weyl solution in
analogy with the spherical symmetry case.

In \cite{letelier} the author shows that the series
\begin{equation}
 \Psi=-\sum_{k=0}^\infty q_k Q_k(x)P_k(y) \
\label{psiprola}
\end{equation}
can be interpreted  as the infinite sum of potentials associated with bars of
equal length $2\sigma$ and linear densities
 $\lambda_k(z)=\frac 12 q_k P_k(z/\sigma)$.  The expression (\ref{psiprola})  is
the metric function, written in prolate spheroidal  coordinates $\{x,y\}$
(see equation (20.6) in page 305 of \cite{stephani} for the definition of these
coordinates) corresponding to the  ERQ
representation
\cite{erq} of the axisymmetric static
vacuum equations, where
$P_k(y)$ and $Q_k(x)$ stand for the Legendre
polynomials and associated Legendre functions of the second kind respectively. It is
clear, from the linearity of the Laplace equation, that this conclusion
concerning the infinite sum of axial bars generating the potential (\ref{psiprola})
is equivalent to the statement that it can  also be generated by a single bar
with linear density:
\begin{equation}
 \lambda(z)=\frac 12\sum_{k=0}^\infty q_k P_k(z/\sigma) \ ,
\label{densilete}
\end{equation}
where $q_k$ are the coefficients of the ERQ representation corresponding to
a particular solution
(\ref{psiprola}).
In fact, the result obtained by Letelier  can be
recovered by means of   considerations other than those used in
\cite{letelier}. We only need to use the relation between the sets of
coefficients of both the Weyl and the ERQ representations  \cite{mq}, \cite{tesis}, concluding that equations
(\ref{phiNG}) and (\ref{mNG})
 mean  that the
Newtonian multipole moments of the bar with its
density given by $\lambda(z)$ (\ref{densilete}) are precisely the coefficients
$a_{2n}$ of the Weyl family of solutions ($M_n^{NG}=-a_n$), 
and the Newtonian gravitational potential is given by the Weyl 
series (\ref{psi}). The detailed calculations are addressed in the Appendix.

Nevertheless, this linear density (\ref{densilete})  is
useless in the general case and, for several reasons,  inefficient for describing the mass distribution
of
the
desired Newtonian objects. First, we point out that except
for the cases where the set of ERQ coefficients $\{q_k\}$  
 of the solution
is finite, the expression for that linear density  cannot be summed explicitly.
The cases of the Schwarzschild solution, defined by $q_0=1$, $q_k=0$,
$\forall k>1$,
as well as  the Erez-Rosen solution \cite{erez}, which only has two
coefficients $q_k$ different from zero
($q_0=1$, $q_2\neq 0$, $q_k=0$, $\forall k>2$)
are,  of course, good examples of this. Second, the definition of the density
(\ref{densilete})  associated with a wide class
of solutions of the Weyl family (all the reflection-symmetric solutions for
instance)
shows a specific anomaly at  both ends of the bar ($t=\pm 1, z=\pm \sigma=\pm
M$) where the
value of the density diverges unless the series of the coefficients $\{q_n\}$
itself converges:
\begin{equation}
 \lambda(z=\pm \sigma)=\frac 12\sum_{k=0}^{\infty}q_{2k} \ .
\end{equation}

Since the aim of the work is to extend the construction of an object with a
representative density
to a wide range of solutions, the establishment of a
relation between the multipole structure
of the Weyl solution and the density of the object  would be a very relevant
success. This achievement would allow us to describe how
the corrections to the spherical symmetry of the Weyl solution are reflected in
a non-constant density of the mass distribution. In our opinion, the efforts of
such research should be focused
on  identifying  the so-called Pure
 Multipole Solutions \cite{mq1}, \cite{mq}, \cite{11} because these
solutions attempt to describe
deviations from spherical symmetry
by means of their Multipole Moments.

In the next section we introduce  a new family of static and axisymmetric
solutions of the Einstein vacuum field equations with a specifically known
multipolar structure, and we derive for it  an interpretation of its metric
function $\Psi$ by means
of the Newtonian gravitational  potential of a particular mass distribution.
This physical object with its characteristics (like its density)
is a sort of mathematical artifice\footnote{Let us note that  expression (\ref{phiNG}),
obtained  from  equation (\ref{mudelta}), is identical
to that derived from the method of singular sources (see \cite{gutsu} and
references therein) used to construct relativistic static solutions with a
singular source.} used to describe the relativistic solution and
it should not be confused with the real non-spherical isolated source that generates
the exterior gravitational field.

\section{Solutions of the Weyl family with a prescribed relativistic multipole
structure}

\subsection{The solution}

In \cite{mq},  the  solution of the Weyl family that
represents the exterior gravitational field of a mass distribution whose
multipole structure
 only possesses mass $M$ and quadrupole moment $Q$ was shown. The procedure for
constructing
this M-Q solution requires knowledge of the relativistic multipole moments (RMM) of a
generic
Weyl solution; that is to say, we need to calculate the relationship between
the
RMM and the coefficients $\{a_n\}$ of the Weyl family of solutions. This
calculation is
done by means of the FHP-method \cite{fhp}
and the coefficients $\{a_n\}$ obtained can be constrained to satisfy the
conditions
imposed on the RMM claimed for the solution. The M-Q solution has become
 a useful tool for describing small deviations from the spherically symmetric
solution \cite{luisgyr}, \cite{luisgeod}. The assumption considered for  the
M-Q solution is that $Q$ is small since we want to approach  the
Schwarzschild solution as closely as possible, and all the RMM of higher order are
negligible\footnote{This consideration about the RMM higher than $Q$ is
supported by the
following argument: The Newtonian calculation
of the multipole moments of an ellipsoidal mass distribution
 leads to  the conclusion that as they increase
 their order they decrease in magnitude proportionally to the powers of the
eccentricity of the ellipsoidal configuration (see \cite{tesis} for details).}.
Thus, the M-Q solution is constructed as a
 sum of functions in a power series of the dimensionless quadrupole parameter
$q\equiv Q/M^3$ starting at the Schwarzschild solution as the first order, in
such a way that  the successive powers of $q$ control the desired corrections
to the spherical symmetry.

We now wish to introduce a similar but different proposal for constructing
relativistic gravitational fields close to the Schwarzschild solution. When
attempting  to describe an isolated compact body that is not spherically
symmetric,
 all the RMM appear no matter how small the deviation is from  being spherical.
Therefore, let us assume that all RMM appear in the solution that we want to
construct but let us restrict their magnitudes to being very small, in such a way
that  we
can neglect all terms in the Weyl coefficients wherever a product of RMM is
involved. This explains the name of the family of solutions: {\it
Linearized Multipole} (LM) solution.
The structure of the Weyl coefficients for that solution is as
follows\footnote{The case of the solution M-Q$^{(1)}$ \cite{mq1}
can be recovered with
the first two terms ($g=1$). The function $F_1(n)$ reduces to the corresponding
expression in \cite{mq1}. The explicit expression (\ref{efe}) is deduced
from the calculation of the RMM of a generic static and axisymmetric solution by
applying the FHP method.
}:
\begin{equation}
 a_{2n}^{LM}=-M^{2n+1} \sum_{k=0}^{g}m_{2k} F_k(n) \ , a_{2n+1}=0 \ ,
 \label{esto}
\end{equation}
where (equatorial symmetry is assumed) $m_{2k}\equiv {\displaystyle
\frac{M_{2k}}{M^{2k+1}}}$ denotes the
dimensionless parameter associated with the multipole moment $M_{2k}$ of order
$2k$, $g$ being the number of RMM in the solution in addition to the
monopole,
and  the functions $F_k(n)$ for each $2^{2k}-$pole moment are defined by
the expression\footnote{Note that $F_k(n)=0$ for $k>n$.}:
\begin{equation}
 F_k(n)= \left\{
\begin{array}{l}
{\displaystyle I(k)\frac{n! (2n-1)!!}{(n-k)! (2n+2k+1)!!} \left[kn+II(k)\right]}
\quad , \quad  1< k \leq n \\
\\
{\displaystyle \frac{1}{2n+1}} \quad , \quad  k=0
\end{array}
\right.
\label{efe}
\end{equation}
with
\begin{eqnarray}
 I(k)&=&\frac{1}{2^{k-1} k(k+1)} \displaystyle{{{4k+1} \choose {2k}}} \nonumber
\\
II(k)&=&\frac k2 (2k^2+k+1) \ .
\end{eqnarray}
We decompose expression (\ref{efe}) in the following form:
\begin{equation}
 F_k(n)= h(k)+\sum_{j=0}^k\frac{h_j(k)}{2n+2j+1} \ ,
\label{efek}
\end{equation}
 where $h(k)$ and $h_j(k)$ are coefficients depending on the $2^{2k}$-pole
moment. The independent term of this decomposition, $h(k)$, is
easily
determined by the following expression,
 since  the numerator of each $F_k(n)$ is a polynomial of degree $k+1$:
\begin{equation}
 h(k)=k \frac{I(k)}{2^{k+1}} \ , \forall \ k>0 \ ,
\label{hachek}
\end{equation}
whereas the other coefficients $h_j(k)$ must satisfy the following relations:
\begin{equation}
 h_j(k)\prod_{i=0,i\neq j}^k(2n_j+2i+1)=I(k) \left(k n_j+II(k)\right)
\prod_{i=0}^{k-1}(n_j-i) \ , j=0\dots k \ ,
 \label{rela}
\end{equation}
where ${\displaystyle n_j\equiv -\frac{2j+1}{2}}$. An easy,
but cumbersome, calculation of the products appearing in (\ref{rela}) leads to
the following explicit expression for the coefficients $h_j(k)$ ($\forall \ k
\geq j$, since $h_j(k)=0$ for $k<j$):
\begin{equation}
h_j(k)=\frac{1}{2^{4k-1}}(-1)^{k-j-2}\displaystyle{{{4k+1} \choose
{2k}}}\frac{k^2+k/2-j}{(k+1)}\frac{(2k+2j)!}{(2j)!(k+j)!(k-j)!} \ .
\end{equation}
With respect to the case of the function $F_0(n)$ we obviously have $h(0)=0$ and
$h_0(0)=1$.

\subsection{Characterization of the solution}

\vspace{3mm}

\noindent {\bf A) The density of the bar.}

In analogy with the homogeneous axial rod of the Schwarzschild solution, we
wish
to introduce a non-constant density on an axial rod whose Newtonian
gravitational
potential reproduces the metric function of the LM solution, and whose Newtonian
multipole moments lead to the Weyl coefficients of the solution.
These coefficients (\ref{esto})
$\{a_n\}$ of the LM solution  are given by the following expression
\begin{equation}
a_{2n}^{LM}=-\sum_{j=0}^g\frac{M^{2n+2j+1}}{2n+2j+1} \frac{H_j}{M^{2j}} -
M^{2n+1} H \ ,
\label{anLM}
\end{equation}
where we have made use of  the decomposition (\ref{efek})  of $F_k(n)$, and the
following definitions:
\begin{equation}
 H\equiv \sum_{k=0}^gm_{2k}h(k) \qquad , \qquad H_j\equiv
\sum_{k=j}^gm_{2k}h_j(k) \ .
\label{haches}
\end{equation}

It is easy to see that if we take $\mu(\hat z)=c_j{\hat z}^{2j}$ for an
arbitrary constant $c_j$, then the corresponding Newtonian moments from
 equation (\ref{mNG}), with the notation $\hat z \equiv XL$, come out as
\begin{equation}
 M_{2n}^{NG}=L^{2n+1}\int_{-1}^1 X^{2n} \mu(LX)dX =\frac{2}{2n+2j+1} L^{2n+2j+1}
c_j\ .
\label{mNGparcial1}
\end{equation}
and the choice $\mu(\hat z)=c L\delta(\hat z\pm L)$
leads to
\begin{equation}
 M_{2n}^{NG}= L^{2n+1} c\ .
\label{mNGparcial2}
\end{equation}
Therefore, the condition $a_{2n}=-M_{2n}^{NG}$ for the coefficients  (\ref{anLM}) is  automatically satisfied  if
the physical object that represents the LM solution is, let us
say, a {\it dumbbell} consisting of two point-like masses of magnitude
 ${\displaystyle \frac{H M}{2}}$
located at both ends of a bar\footnote{These masses arise from a function $\mu(X)=\frac 12 H  \left[
\delta(X+1)+\delta(X-1)\right]$ (see equation (\ref{mNGparcial2})) considered in addition to the  linear density of
the bar of the dumbbell. An alternative argument comes from the
direct comparison
of the gravitational potential of two particles of mass $m$ situated at
distances $-L$ and $L$ respectively along the $Z$ axis, ${\displaystyle \Psi
=\Phi=
-\frac{m}{\sqrt{\rho^2+(z+L)^2}}-\frac{m}{\sqrt{\rho^2+(z-L)^2}}}$, and the
corresponding metric function by means of performing a power
series expansion as follows: ${\displaystyle \Phi=
-\frac{m}{\sqrt{\rho^2+(z+L)^2}}-\frac{m}{\sqrt{\rho^2+(z-L)^2}}=
-\sum_{n=0}^{\infty}\frac{P_{2n}(\omega)}{r^{2n+1}} \left(2m L^{2n}\right)}$,
and therefore the Weyl coefficients related to this potential of two particles
are $a_{2n}=-2m L^{2n}$. Let us note that
${\displaystyle \frac{1}{\sqrt{\rho^2+(z-x)^2}}=\frac 1r
\sum_{n=0}^{\infty}\left(\frac
xr\right)^n P_n(z/r)}$.} of half-length $L=M$, whose linear density is
constructed with the coefficients $H_j$ (\ref{haches}) as follows:
\begin{equation}
 \mu(X)=\frac 12 \sum_{j=0}^g X^{2j} H_j  \qquad , \ X \in [-1,1] \ .
\label{densiX}
\end{equation}

Before continuing to analyze the properties of the linear density of the bar $\mu(X)$, the
following must be emphasized again: We claim to represent a
certain type of Weyl solutions
by means of the Newtonian gravitational potential  of
a physical object that resembles  a dumbbell, whose characteristics are  two
identical masses  and the non-constant density of
the bar connecting both of them. It is true that this paradigm is only a
mathematical convenience, and the dumbbell must not be regarded as representing
a real
 physical mass distribution lying along the $Z$ axis of the space-time.
Nevertheless, the implementation of this representation is, as we
 shall see in the following  sections, a useful tool for describing the physical
characteristics of  a static and axisymmetric space-time.

First, with this picture in mind we generalize the known feature of  the Schwarz\-schild
solution to the LM solution that represents  a relativistic Weyl solution
 whose first $g$  RMM, in addition to  the
monopole moment, are non-zero. The difference with respect to the Schwarzschild solution
consists of the change in the density of the bar and the addition of
masses.
Within this representation no divergence such as that appearing in the definition
of the density made by Letelier
\cite{letelier} occurs, and the dumbbell admits a well-behaved linear
density for its bar.

Second,  the following are very important points of this result:
on the one hand, the density (\ref{densiX}) completely
defines  the whole structure of the dumbbell since the additional masses can be
calculated
 from it, as we shall see in the next section. On the other hand, this density
is
constructed in terms of the RMM of the source and hence it provides
a mechanism for adjusting the contributions of different RMM to the changing aspect of
the physical object that represents the non-spherical behaviour of the
relativistic solution.

In the following section we shall investigate the properties of the density
(\ref{densiX}) in detail and the interpretation of the
 masses of the dumbbell. In section $4$ we implement some examples and provide
relevant
conclusions from comparisons among the different cases.

\vspace{5mm}

\noindent {\bf B) Properties and characteristics of the dumbbell}.

\vspace{2mm}

{\it i)} The density of the bar at its ends  equals the value of the density
for
the spherical case (Schwarzschild), i.e., $\mu(X=\pm1)=1/2$ since the following
relation
is fulfilled:
\begin{equation}
 \sum_{j=0}^gH_j = 1 \ .
\label{sumaH}
\end{equation}
It is easy to see that (\ref{sumaH}) holds because the structure of the
functions $F_k(n)$ is such that  the sum of all the coefficients $h_j(k)$ for
each
 $k$ (multipole order) vanishes:
\begin{equation}
 \sum_{j=0}^g h_j(k) =0 \qquad , \  k=0 \dots  g \ .
\label{relaindep}
\end{equation}

\vspace{2mm}

{\it ii)} The masses of the dumbbell have been calculated  previously as
$MH/2$. Since we explicitly
know the value of $h(k)$ for each multipole moment (\ref{hachek}), we
can rewrite and evaluate the mass of each ball as follows:
\begin{equation}
 {\displaystyle H= \sum_{k=0}^g \frac{m_{2k}}{2^{2k} (k+1)} {{4k+1}
\choose {2k}} } \ .
\label{masabolab}
\end{equation}
Alternatively, we can calculate the mass of the balls by means of the excess or
deficit (depending on the sign of the RMM)
provided by the density of the bar with respect to the total mass $M$ of the
solution. The Newtonian  zero-order multipole moment  of the bar is given by
the following expression ($L\equiv M$):
\begin{equation}
 a_0^{bar}=- \int_{-L}^L \mu(z) dz=-2 L \int_0^1 \mu(X)
dX=-M\sum_{j=0}^g\frac{H_j}{2j+1} \ ,
\end{equation}
and it represents the mass of the bar except for the sign. Therefore, the
following relation must hold for the total mass of the dumbbell:
\begin{equation}
 -a_0^{bar}+2 \nu = M \ ,
\end{equation}
where $\nu$ denotes  the mass of each ball of the dumbbell, and hence we have
\begin{equation}
 \nu=\frac M2\left[1-\sum_{j=0}^g\frac{H_j}{2j+1}\right] \ .
\label{masabolanu}
\end{equation}
In fact, this expression is true, i.e., we can prove that $\nu=MH/2$ from the
structure of the functions $F_k(n)$:
the equivalence between $\nu$ (\ref{masabolanu}) and $MH/2$ (\ref{masabolab})
leads to
\begin{equation}
 1-\sum_{j=0}^g\frac{1}{2j+1}\sum_{k=j}^gh_j(k)m_{2k}=\sum_{k=0}^g h(k) m_{2k} \
,
\end{equation}
or equivalently, for each multipole order $k>0$ we have that ($h_0(0)=1$ and
$h(0)=0$ since
$m_0=1$)
\begin{equation}
 \sum_{j=0}^g\frac{h_j(k)}{2j+1}=-h(k) \quad , \quad \forall k=1 \dots g \ ,
\end{equation}
a relation that is fulfilled for every $k$ since $F_k(n=0)=0$.

In conclusion, the dumbbell representing the LM solution is characterized
exclusively  by the linear density of its bar (\ref{densiX}). The mass of  each
ball of the dumbbell is given by $\nu$ (\ref{masabolanu}). Accordingly, the
set of coefficients $\{H_j\}$ determines the Weyl solution and characterizes it
 for the  $2^{2g}$-pole order considered. We work with $g+1$  coefficients but
owing to the constraint, (\ref{sumaH}), we only have $g$
independent coefficients $H_j$,
and so we construct the LM solution having $g$ multipole moments in
addition to the monopole with $g$ independent coefficients in accordance with the
$g$ dimensionless multipole moments $m_{2k}$.

\vspace{5mm}

\noindent {\bf C) The metric function as a gravitational potential.}

The interpretation of the LM solution by means of the gravitational potential
of the dumbbell allows us to calculate the metric function of the solution as
follows
\begin{equation}
 \Psi=\Phi=\int_{-1}^1\frac{-M \mu(X)}{\sqrt{\rho^2+(z-MX)^2}} dX -
\frac{\nu}{\sqrt{\rho^2+(z-M)^2}}-\frac{\nu}{\sqrt{\rho^2+(z+M)^2}},
\label{66}
\end{equation}
with $\mu(X)$ the linear density  (\ref{densiX}) of the bar of the dumbbell  and $\nu$
being
the mass of each ball (\ref{masabolanu}).
Since the density $\mu(X)$ is a polynomial we have to evaluate the integrals of  the following
 type:
\begin{equation}
 \Upsilon_{\alpha}\equiv \int_{-1}^1\frac{X^{\alpha}}{\sqrt{\rho^2+(z-MX)^2}} dX
\end{equation}
using the recurrence relation
\begin{equation}
 \Upsilon_{\alpha}=\Gamma_{\alpha}(X)\sqrt{\rho^2+(z-MX)^2}
|_{-1}^1+\Lambda_{\alpha} \Upsilon_0 \quad , \quad \alpha \geq 1 \ ,
\end{equation}
where ${\displaystyle \Gamma_{\alpha}(X)\equiv
\sum_{i=0}^{\alpha-1}p_i(\alpha)X^i}$ is a
polynomial of degree $\alpha -1$ and $\Lambda_{\alpha}$ a function of the
Weyl coordinates $\{\rho,z\}$ not depending on the variable $X$. A cumbersome
calculation leads to the following iterative scheme ($r\equiv\sqrt{\rho^2+z^2}$
denotes the radial Weyl
coordinate):
\begin{eqnarray}
 & &{\displaystyle \Lambda_{\alpha}=M z p_0(\alpha) -r^2 p_1(\alpha) }\\
\nonumber
& & {\displaystyle  p_{\alpha-1}(\alpha)=\frac{1}{\alpha M^2} \qquad , \qquad
p_{\alpha-2}(\alpha)=\frac{z (2\alpha-1)}{\alpha (\alpha-1)M^3}} \\ \nonumber
& &{\displaystyle 0 =
r^2(\alpha-k)p_{\alpha-k}(\alpha)-z M(2\alpha-2k-1)p_{\alpha-(k+1)}(\alpha)+}\\
\nonumber
& &{\displaystyle +M^2(\alpha-k-1)p_{\alpha-(k+2)}(\alpha) \qquad , \  \forall \
k\geq 1, \ \alpha\geq 1 \ ,}
\label{eqQ}
\end{eqnarray}
and the integral
\begin{equation}
 \Upsilon_0=\frac 1M \ln\left(\frac{z-M-r_{-}}{z+M-r_{+}}\right) \quad , \quad
r_{\pm}\equiv\sqrt{\rho^2+(z\pm M)^2} \ .
\label{upsi0}
\end{equation}
Finally, the metric function is given by the following expression, where we
obviously recover the Schwarzschild solution for the case $g=0$:
\begin{eqnarray}
 \Psi=\Phi&=&-\frac M2 \sum_{j=0}^g\int_{-1}^1\frac{X^{2j}
H_j}{\sqrt{\rho^2+(z-MX)^2}} dX +\nonumber \\
&-&\frac{\nu}{\sqrt{\rho^2+(z-M)^2}}-\frac{\nu}{\sqrt{\rho^2+(z+M)^2}} =
\nonumber \\
&=&\frac 12 \left[H_0+\sum_{j=1}^gH_j\Lambda_{2j}\right]
\ln\left(\frac{z+M-r_{+}}{z-M-r_{-}}\right)+  \\
&+&\frac M2\left[\sum_{j=1}^gH_j\Gamma_{2j}(-1)\right]r_+-
\frac
M2\left[\sum_{j=1}^gH_j\Gamma_{2j}(1)\right]r_--\nu\left[\frac{1}{r_+}+\frac{1}{
r_-}\right] \nonumber
\label{psiLM2}
\end{eqnarray}

\section{Some examples and constraints to the multipole moments}

\subsection{The Erez-Rosen {\it vs}  the M-Q$^{(1)}$ solution}

These are  both two-parameter solutions of the Weyl family, one of the parameters
($M$)
representing the mass and the other $q$ or $q_2$ denoting the
dimensionless quadrupole moment for the M-Q$^{(1)}$ or the ER solution
respectively. In
\cite{luisgyr} a comparison between
these solutions was developed and different conclusions  were obtained
regarding the
behaviour  of a gyroscope precessing in circular orbits in each
 space-time. The Weyl coefficients of the ER solution are known
\cite{tesis}:
\begin{equation}
 a_{2n}^{ER}=-\frac{M^{2n+1}}{2n+1}\left(1+q_2\frac{2n}{2n+3}\right)=-M^{2n+1}
\left[\frac{1-q_2/2}{2n+1}+\frac{3q_2/2}{2n+3}\right] \ .
 \label{anER}
\end{equation}
From this expression for the Weyl coefficients it immediately follows that  the ER
solution can be represented by the Newtonian gravitational potential
of a bar with linear density:
\begin{equation}
 \mu^{ER}(X)=\frac 12 \left(1-\frac{q_2}{2}\right)+\frac 34 q_2 X^2 \ ,
 \label{densiER}
\end{equation}
since the multipole Newtonian moments (\ref{mNG}) of a bar with this linear
density reproduce  equation (\ref{anER}).
As can be seen, this result recovers (\ref{densilete}) Letelier's definition
\cite{letelier}
for
this solution  because the ER solution has a finite number of
coefficients $\{q_n\}$ in the ER representation.
 From the density $\mu^{ER}(X)$ we can also obtain
  the  Newtonian
potential (\ref{phiNG}),
\begin{eqnarray}
\Phi=\Psi&=&-\frac 12 \left[1-\frac
{q_2}{2}+\frac{3q_2}{4M^2}\left(2z^2-\rho^2\right)\right]\ln\left(\frac{
M-z+r_{-}}{-M-z+r_{+}}\right)+\nonumber\\
&-&\frac{3q_2}{8M^2}\left[(3z+M)r_{-}-(3z-M)r_{+}\right] \ .
\label{psiER}
\end{eqnarray}
A  straightforward calculation allows us to prove that this
expression (\ref{psiER}) reproduces the known metric function of the ER
solution in prolate spheroidal coordinates (see equation (20.9) on page 306 of
\cite{stephani} for details.)

\vspace{2mm}

In contrast to the ER solution, the M-Q$^{(1)}$ solution entails a different
representation. This solution is the linear order in the quadrupole parameter of
the
solution M-Q \cite{mq}. The Weyl coefficients of this
solution are known \cite{mq}-\cite{sueco}:
\begin{equation}
 a_{2n}=-\frac{M^{2n+1}}{2n+1}\left(1+q\frac{5n(n+2)}{2n+3}\right)
\label{aesmq1}
\end{equation}
where the parameter $q\equiv{\displaystyle \frac{Q}{M^3}}$ is used to denote the
dimensionless quadrupole moment. As we have previously noted,
 the sum of the series for the Letelier density (\ref{densilete}) is not
available is this case, since the set of ERQ coefficients $\{q_n\}$
of this solution is not finite (see \cite{mq1} for details): $ q_0=1,
q_2=\frac{15}{2} q, q_{2n}=\frac 54 (4n+1)q
\ , \forall n\geq 2$. From  equation (\ref{mNG}) and the Weyl coefficients
(\ref{aesmq1}) it is easy to see that the M-Q$^{(1)}$ solution can be
represented by a
dumbbell consisting of a bar of length $2L$ with linear density\footnote{The
Weyl
coefficients  can
be decomposed as follows: ${\displaystyle
a_{2n}=-\frac{M^{2n+1}}{2n+1}\left(1-\frac{15q}{8}\right)-\frac{M^{2n+3}}{2n+3}
\left(\frac{15q}{8M^2}\right)-M^{2n+1}\frac 54 q}$.}
\begin{equation}
\mu^{M-Q^{(1)}}(X)=\frac 12 \left(1-\frac{15}{8}q\right)+\frac{15}{16}q X^2
\end{equation}
and a point-like particle  at each  end of the dumbbell with
respective masses ${\displaystyle \nu=\frac 58 qM}$.

We now  compare both solutions, analyzing the corresponding linear
densities of their bars, which are polynomials of the form $\mu(X)=A+BX^2$. The
maximum or the minimum (depending on the sign of the
quadrupole parameter, $q,q_2<0$ or $q,q_2>0$ respectively) of the density  is
located at the origin and is given by $\mu(0)=A$, as can be seen in  Figure 1.

\begin{figure}[ht]
$$
 \begin{array}{ll}
\epsfig{figure=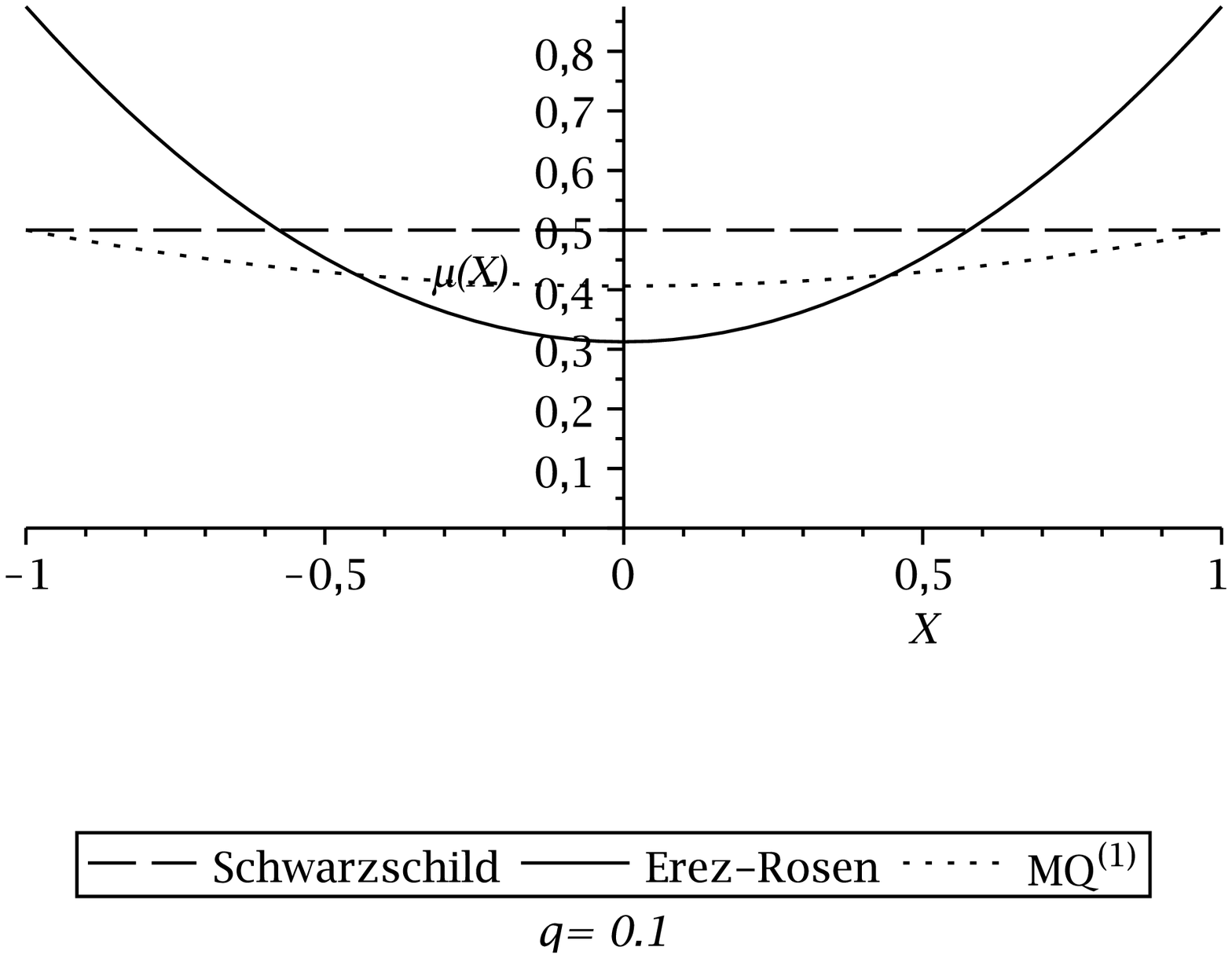,height=2.6in}
&
\epsfig{figure=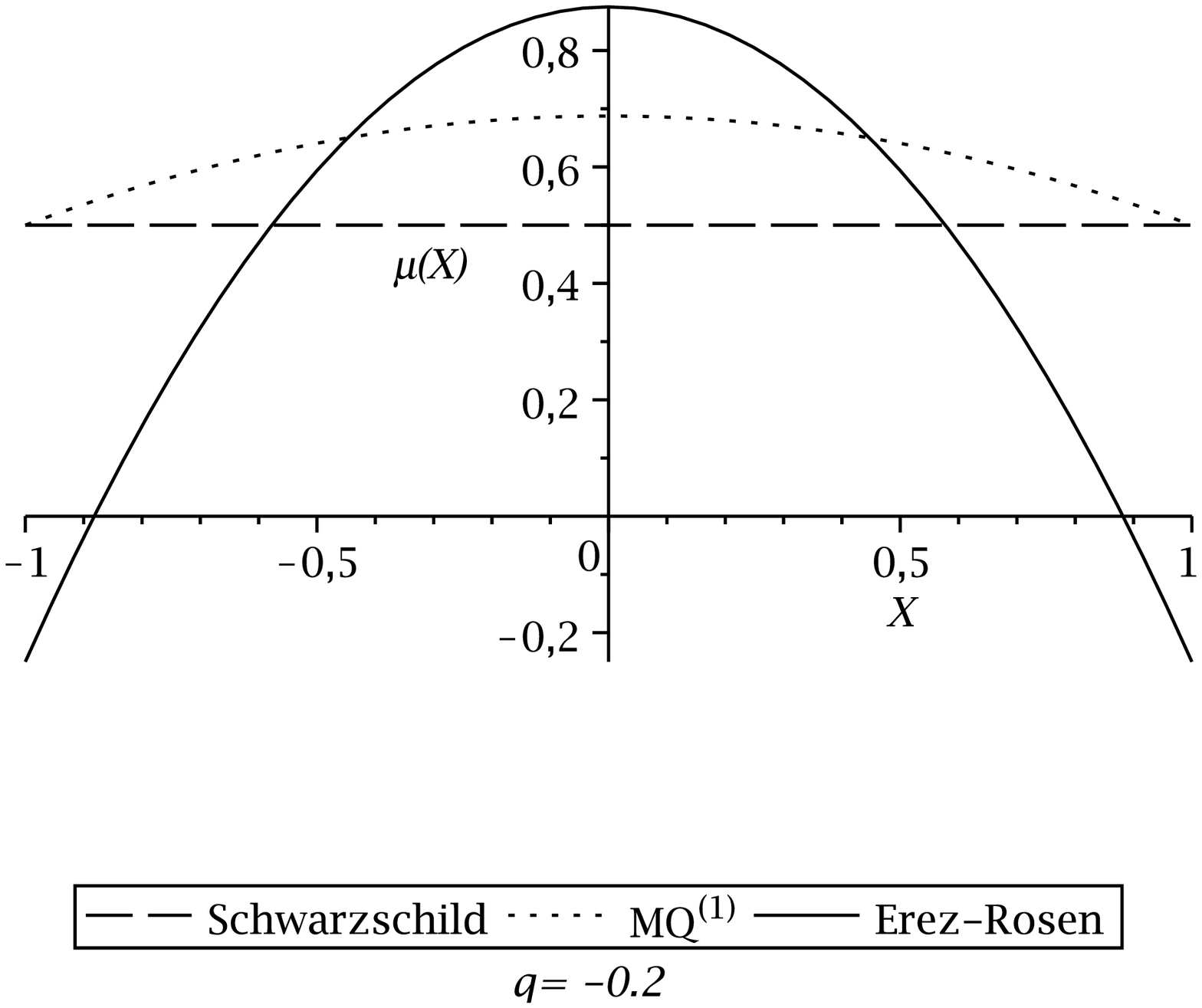,height=2.4in}\nonumber
\end{array}
\nonumber
$$
\caption{\it Linear density profiles for different values of the parameter $q$
corresponding to the M-Q$^{(1)}$ and Erez-Rosen solution compared with the
Schwarzschild constant linear density. The positive sign of $q$  provides a
density representing an elongated object whereas $q<0$ represents a flattened
object, which agrees with a decrease or increase in the density at the center of
the bar respectively.}
\end{figure}

The following remarks can be made about the behaviour of the density.
The ER solution is represented by a bar with linear density
(\ref{densiER}) rather than a dumbbell, as is the case for the M-Q$^{(1)}$
solution. Let us note that we can evaluate the total mass of the solution by
calculating the Weyl coefficient $a_0$ as the integral of the density along the
bar, and we obviously obtain  $a_0=-M$.

 The density of the M-Q$^{(1)}$ solution is equal to $1/2$ at
both ends of the bar, as shown in Figure $1$. Therefore, for $q<0$ the linear
density is positive definite along the whole bar and it has the maximum value
at $z=0$: $\mu(X=0)=A=\frac 12 (1-\frac{15}{8}q)$, whereas for $q>0$ a suitable
definition of the density requires the following constraint:
\begin{equation}
A\geq 0 \Rightarrow q\leq \frac{8}{15} \ .
\end{equation}

 The ER solution shows slight differences with respect to such
behaviour: for $q_2<0$ we require $\mu(X=\pm1)\geq0$ , which leads\footnote{Or,
equivalently, we demand the roots of the polynomial to be greater than $1$ (in
absolute value): $|\sqrt{-A/B}|\geq 1$.} to the following constraint on the
quadrupole parameter:
\begin{equation}
\frac{-1+q_2/2}{3/2q_2} \geq 1 \Leftrightarrow  |q_2| \leq 1
\end{equation}
whereas for the case $q_2>0$ we need $A\geq 0 \Leftrightarrow q_2\leq 2$.

In conclusion, the linear density of the M-Q$^{(1)}$ solution is positive
definite for the range $q \in \left(-\infty,\frac{8}{15}\right]$ whereas the
ER solution is characterized by a linear density that is positive definite for
the range of the quadrupole parameter $q_2 \in [-1,2]$. Since the relation
between $q_2$ and $q$ is known, because they both represent the dimensionless
quadrupole moment ($q_2=\frac{15}{2} q$), the quadrupole moment $Q$ is
constrained as follows:
\begin{equation}
Q \in \left\{
\begin{array}{c}
M^3 \left(-\infty,\frac{8}{15}\right] \ , \qquad M-Q^{(1)}\\
\\
M^3 \left[-\frac{2}{15}, \frac{4}{15}\right] \ , \qquad ER
\end{array}
\right.
\label{range}
\end{equation}
The upper limit of $Q$ for the M-Q$^{(1)}$ solution is
exactly twice the magnitude of that of the ER solution and it does not have  a lower
bound, whereas the
quadrupole moment must be higher than a minimum value for the ER
solution.

As we already noted in the previous section, this representation of the Weyl
solutions by means of the artificial construction of a dumbbell allows
 us to characterize the properties of the  Weyl solution. Let us go deeper into
this assessment by making  an interpretation of the
behaviour of the  linear densities we have obtained.

\vspace{3mm}

{\noindent i)} In a vacuum relativistic solution,  a  negative value of the
quadrupole parameter $q$ is
associated with a flattened object whereas $q>0$ denotes an elongated source. With
the representation of  the Weyl solution by means of a dumbbell
this characteristic of the source can be modeled by the behaviour  of the bar
density as follows: if
 $q>0$ the linear density of the bar shows a deficit of
mass with respect to the constant linear density $\mu=1/2$ that corresponds to
spherical symmetry (see Figure 1), at the same time as  the masses of the
point-like particles (the balls of the dumbbell) are
positive. In the opposite case, for $q<0$ the linear density shows an
excess of mass with respect to $\mu=1/2$, which is related to a negative value for
the masses of the balls (deficit of mass at the ends of the bar).
This behaviour of the density is therefore in concordance with the model
picture of an
elongated/flattened object in comparison with the spherically symmetric
configuration.

\vspace{3mm}

{\noindent ii)} The ranges of values (\ref{range}) obtained for $Q$ in both
solutions  are consistent since our family of LM solutions requires
small values for their multipole parameters to maintain the physical interpretation
we have provided for such solutions. However, these ranges should not be interpreted as bounds on the
real values
 of the multipole moments that non-spherical bodies of anisotropic fluids,
eventually considered as sources of these Weyl exterior solutions, could
have.
If the density is not positive definite this means that the method used
to represent the potential $\Psi$ breaks down for values outside
the ranges obtained.

\vspace{3mm}

{\noindent iii)} Nevertheless, some interpretation can be attempted regarding these ranges
of values. It is known (see \cite{forstchrite} for details) that the horizon of
the Erez-Rosen metric, i.e., the hypersuface $x=1$
(where $x$ denotes the radial prolate coordinate, or equivalently $r=2M$ in
Schwarzschild coordinates) can be divided into two parts: the first part represents
the horizon (where the norm of the time-like Killing vector is null) and the second
corresponds to the Killing singularity, where that vector becomes infinite. For
instance, if $0\leq q_2\leq 2$ the horizon totally covers the hypersurface
$x=1$, but
 for other values of the quadrupole parameter $q_2$ there is only a restricted range
of values for the angular coordinate that makes $g_{00}^{ER}=0$ on the surface
$x=1$. A detailed
 study of the infinite redshift surface $g_{00}=0$ was done in \cite{tesis} (see
web address to download file) with respect to the M-Q$^{(1)}$ solution. There it was
established that there is a range of values for the quadrupole parameter $q$  where the
horizon totally covers the hypersurface $x=1$. For values of $q$ outside this
range
 that surface  diverges for certain angular regions. The
range of
values of $q$ is exactly that obtained  when we require the density
of the dumbbell
to be positive definite (\ref{range}).

Thus, there is another feature of the dumbbell representation: a 
positive or negative value  of the density allows us to characterize whether the
surface
$g_{00}=0$ completely covers the surface $x=1$ or not.
 If $\mu(\hat z)>0$ (values of $q$ within the range),  then $e^{2\Psi}\equiv
g_{00}=0$ for all values of the angular coordinate, whereas for $\mu(\hat z)<0$
(values of $q$ outside the range) that surface diverges. The behaviour of the
metric function $\Psi$ for different values of the density can be
interpreted in that sense\footnote{Note that the relation between the Weyl radial
coordinate $R$ and the Schwarzschild coordinate $r$ or prolate  radial
coordinate $x$ is ${\displaystyle R=\sqrt{r^2+ M^2 \cos^2 \theta-2 M r} =M
\sqrt{x^2+y^2-1}}$. Hence, the Schwarzschild surface $r=2M$ ($x=1$) corresponds
to the rod of the dumbbell since $R=My$ (or equivalently $\rho=0$, $z=My$) and
$-1\leq y \leq 1$.}: if $\rho$ tends to zero with $-M<z<M$,  then $\Psi$
goes to $-\infty$ if the linear density $\mu(\hat z)>0$: i.e., when we are
approaching
the source $g_{00}$ tends to zero if the density is positive (the horizon is
reached). By contrast,  $\Psi$ tends to infinity if
the density $\mu(\hat z)<0$ when approaching the source, and hence $g_{00}$
tends to infinity, because we find the Killing singularity.

In conclusion, the positive-definite condition of the density is tantamount to
guaranteeing that the vacuum solution possess a horizon that covers the
hypersurface $x=1$ completely. Otherwise a naked singularity is found when approaching the
source.

\vspace{3mm}

{\noindent iv)} A final comment on the dumbbell's representation. The Weyl
series (\ref{psi}) can be rewritten as the following linear superposition
of Curzon solutions:
\begin{equation}
{\displaystyle{
\Psi=\sum_{n=0}^{\infty}\frac{a_n}{r^{n+1}}P_n(\cos\theta)=\sum_{n=1}^{\infty}}
\frac{c_n}{\sqrt{\rho^2+(z-x_n)^2}}} \ ,
\label{ocho}
\end{equation}
where the  Curzon solution \cite{curzon}
corresponds to  a point-like particle with mass $c_n$ located at the point $x_n$
on the $Z$ axis. Equation (\ref{ocho}) for the metric function $\Psi$
requires the following relation between the Weyl coefficients and the parameters
$c_n$, $x_n$:
\begin{equation}
 a_n=\sum_{x=1}^{\infty} x_i^n c_i  \ .
\label{algcond}
\end{equation}
Well known results in Newtonian potential theory allow us to approximate a
continuous line density with a series of point masses in such a way that the
corresponding
gravitational potential can be written as a linear combination of Curzon
solutions. In fact, it is easy to see that a linear density of the type
$\mu(z)=\sum_i c_i
\delta(z-x_i)$ provides
  a potential $\Psi$  (\ref{ocho}) in such a way that  the relation
(\ref{algcond}) is satisfied from equation (\ref{mNG}), and the continuous limit
of the discrete density
leads to the linear density constructed in the above cases for certain sets of
the discrete parameters $\{x_i\}$, $\{c_i\}$.

\subsection{The Monopole-Quadrupole-$2^4$-pole solution}

Analogously to the method previously used for other cases, the
LM solution can be represented by the gravitational potential of a dumbbell
with the linear density of its bar given by the following expression
(see (\ref{densiX}) and  (\ref{haches}) for details):
\begin{equation}
\mu^{LM}(X)=\frac 12
\left[1-\frac{15}{8}\left(q-\frac{21}{4}m_4\right)+\frac{15}{8}\left(q-42
m_4\right)X^2+\frac{2205}{32}m_4 X^4\right] \ ,
\label{densiLM}
\end{equation}
and the masses of the particles at both ends of the bar are given by
\begin{equation}
\nu=\frac M8 \left(5q+\frac{21}{2} m_4\right) \ ,
\label{masaball}
\end{equation}
This density shows a maximum or a minimum, depending on the sign of $H_1$, at
the origin $\mu(X=0)=\frac 12 H_0$:  if $H_1>0$, or equivalently
if $q-42 m_4<0$,  then the density possess a maximum  and two
minima at both symmetric points ${\displaystyle
X_{min}=\pm\sqrt{\frac{-H_1}{2H_2}}}$, whereas for the case $q-42 m_4>0$ it
is even about $X=0$
but with a minimum at the origin and two maxima. Therefore,  the magnitudes and
the relative signs of the quadrupole and $2^4$-pole moments
determine the slope of the linear density. Since the LM solution
 is constructed under the assumption that all the RMM are equally small
quantities, we shall consider ${\displaystyle
\left|\frac{m_4}{q}\right|=1}$, (
 $|m_4|\approx |q|$) and hence we have the following density for the different
cases:
\begin{equation}
 \mu^{LM}(X)=\left\{
\begin{array}{c}
{\displaystyle \frac 12
\left(1+\frac{255}{32}q-\frac{615}{8}qX^2+\frac{2205}{32}q X^4\right)} \quad , \
m_4/q=1\\
 (q>0 \rightarrow X_{max}=0, q<0\rightarrow X_{min}=0)\\
\\
{\displaystyle \frac 12
\left(1-\frac{375}{32}q+\frac{645}{8}qX^2-\frac{2205}{32}q X^4\right)} \quad , \
m_4/q=-1 \\
 (q>0 \rightarrow X_{min}=0, q<0\rightarrow X_{max}=0)
\end{array}
\right.
\label{densiLMcasos}
\end{equation}

Referring back to point (iii) of the previous section, if we wish to avoid the possibility of a naked singularity we constrain
the linear density to be positive definite, and therefore the
condition $\mu(X_{min})\geq0$ must be fulfilled if $X=0$ represents
 a maximum and
 $\mu(0)\geq0$ if $X=0$ represents a minimum of the linear density, i.e.
the following relations are found:
\begin{equation}
 \begin{array}{cc}
\mu(X_{min})^{LM}\geq 0 \Leftrightarrow H_0-\frac{H_1^2}{4 H_2}\geq 0 &\ ,\
m_4/q=1, q>0  \\
 & \   m_4/q=-1 , q<0\\
\mu(X_{min}=0)^{LM}\geq 0 \Leftrightarrow H_0\geq 0 & \ , \ m_4/q=1, q<0 \\
 & \  m_4/q=-1 , q>0
\end{array}
\label{condLM}
\end{equation}

These conditions, (\ref{condLM}) and (\ref{densiLMcasos}), lead to different
constraints on the magnitude of the parameters
$q$ and $m_4$ depending on their relative sign, which can be summarized as follows
\begin{equation}
 \frac{m_4}{q}=1 : \left\{
\begin{array}{ccc}
 q>0 \ :& \ q\leq \frac{1568}{21125}&\\
& & \qquad \Rightarrow q \in \left[-\frac{32}{255} , \frac{1568}{21125}\right]\\
 q<0 \ :& \ |q|\leq\frac{32}{255}
\end{array}
\right.
\label{constraint1}
\end{equation}
\begin{equation}
 \frac{m_4}{q}=-1 : \left\{
\begin{array}{ccc}
 q>0 \ :& \ q\leq \frac{32}{375}&\\
 & & \qquad \Rightarrow q \in \left[-\frac{1568}{18605} ,
\frac{32}{375}\right]\\
 q<0 \ :& \ |q|\leq\frac{1568}{18605}
\end{array}
\right.
\label{constraintm1}
\end{equation}
Or, in other words, if the quadrupole ($q$) and $2^4$-pole ($m_4$) moments have
equal  sign regardless of whether the object is flattened or elongated,  then the magnitude of both
parameters must be restricted
to the approximate values $[-0.1255,0.0742]$ in order to
conserve a positive definite linear density. In the other cases, $q$ and $m_4$ being
of
different sign, the approximate
constraint $[-0.0843,0.0853]$
is required for a  well behaved linear density.

Figure 2  shows schematic representations of the dumbbell (or a bar
alone) along with their densities and the masses of the balls for
different
 solutions, starting from the spherically symmetric case (Schwarzschild).

Finally we show the gravitational potential calculated explicitly for this
density (\ref{densiLM}). According to expression (\ref{66}) and
taking into account  equations (\ref{psiLM2}), we find
\begin{equation}
 \Phi=\frac 12 \gamma(\rho,z) \ln\left(\frac{z+M-r_{+}}{z-M-r_{-}}\right)+\frac
M2 \left[\beta_{-}(\rho,z)r_{+}-\beta_{+}(\rho,z) r_{-}\right]-
\nu\left(\frac{1}{r_{+}}+\frac{1}{r_{-}}\right),
\label{psiLMrz}
\end{equation}
where
\begin{eqnarray}
 \gamma(\rho,z)&\equiv& H_0+H_1\left(\frac{2z^2-\rho^2}{2M^2}\right)+H_ 2
\left[\frac{1}{8M^4}(8z^4+3\rho^4-24\rho^2z^2)\right]\nonumber\\
\beta_{\pm}(\rho,z)&\equiv&\frac{15}{16}\frac{q}{M^3}(3z\pm
M)-\frac{105}{128}\frac{m_4}{M^3}\left(95 z\pm27 M\right)+\nonumber\\
&+&\frac{2205}{768}\frac{m_4}{M^5}(50z^3-55\rho^2z\pm26Mz^2\mp9M\rho^2) \ .
\end{eqnarray}
$H_i$ are the coefficients of the density (\ref{densiLM}) except for a factor
$1/2$, and $\nu$  represents the mass (\ref{masaball}) of each ball of the
dumbbell, which is
given by (\ref{masabolanu}) or equivalently (\ref{masabolab}). It should be noted that
this expression (\ref{psiLMrz}) leads to the gravitational potential of the
M-Q$^{(1)}$ solution
 for $m_4=0$.

\begin{figure}[ht]
$$
 \begin{array}{cc}
\epsfig{figure=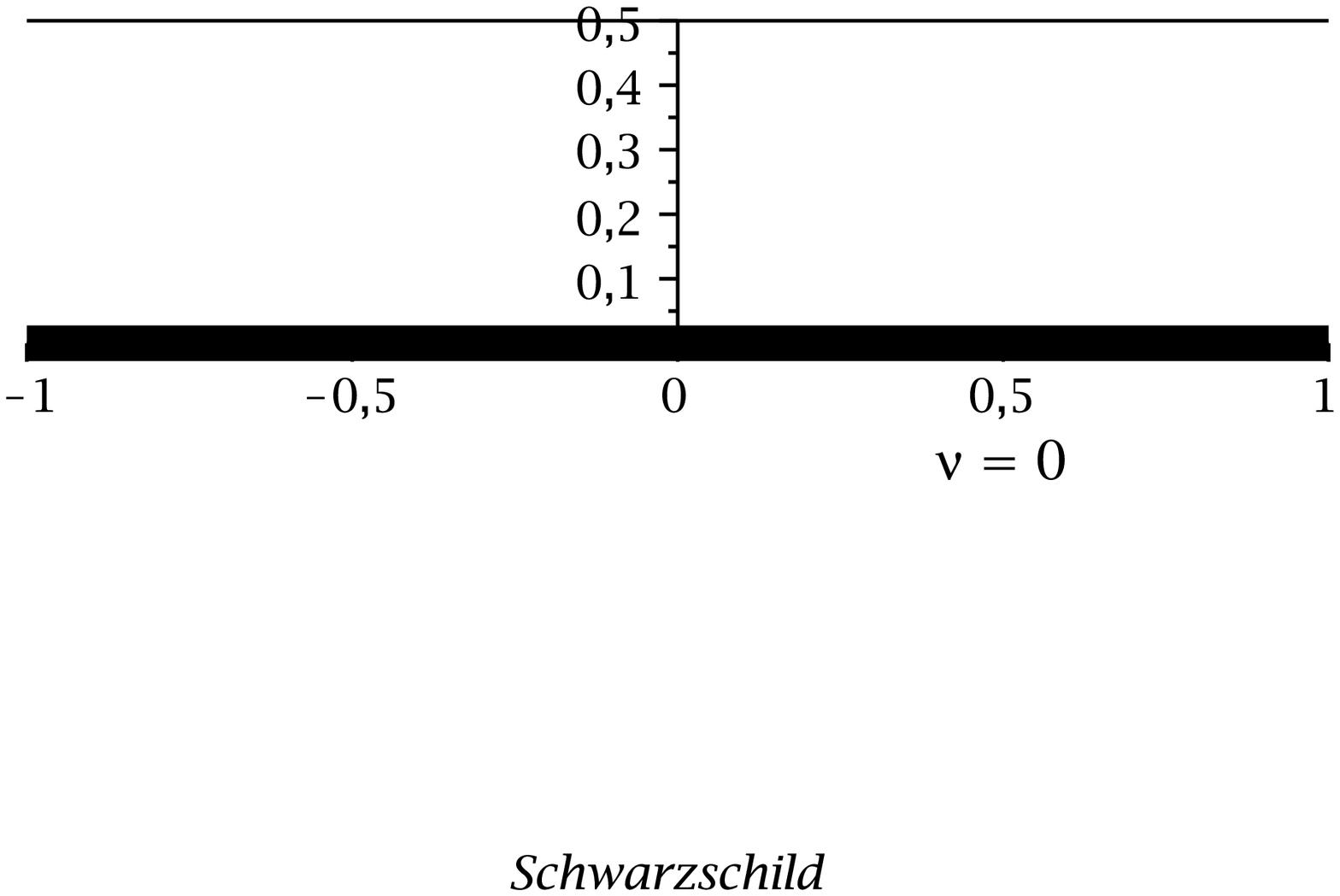,height=2.1in}
&
\epsfig{figure=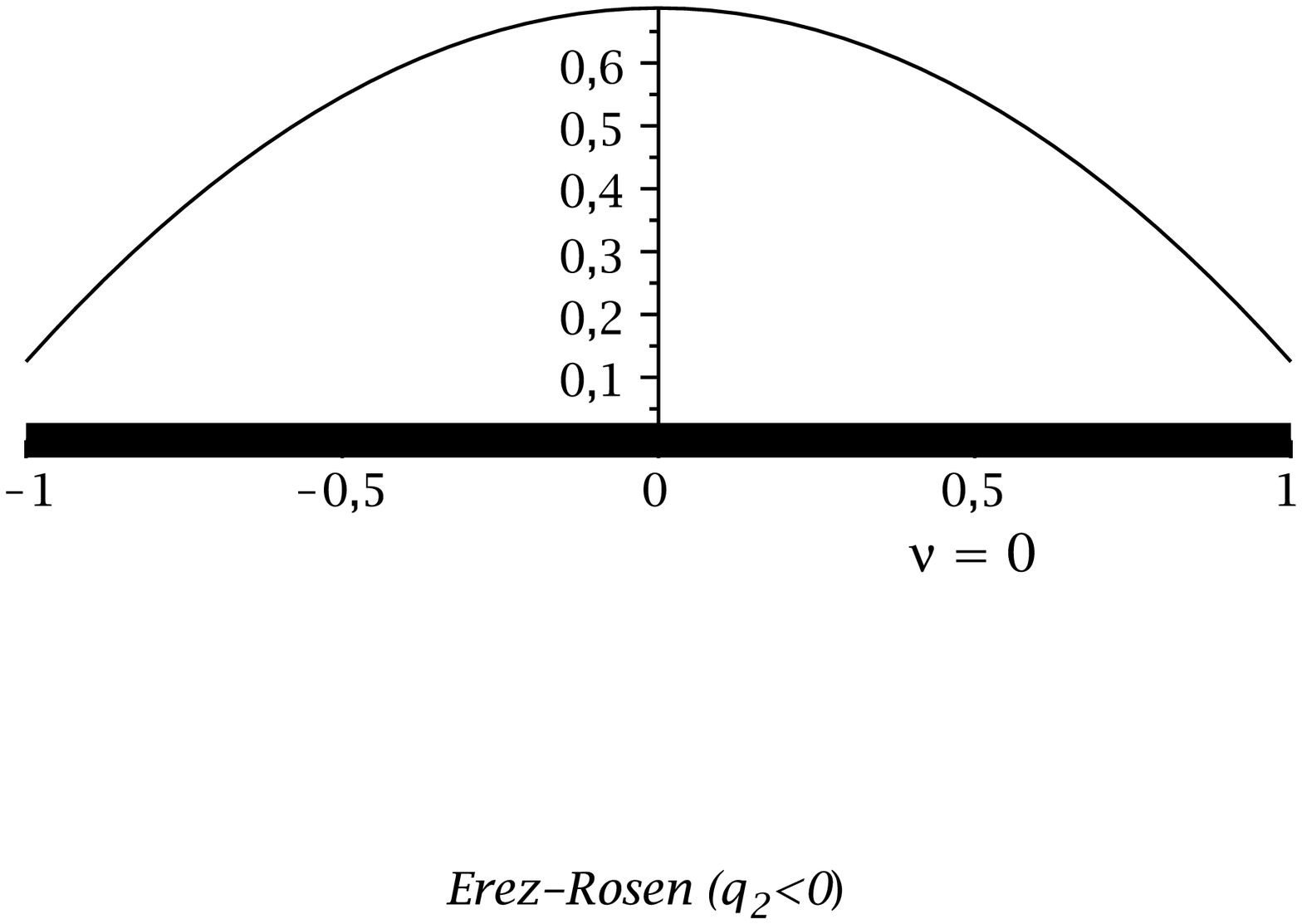,height=2.1in}\nonumber \\
\epsfig{figure=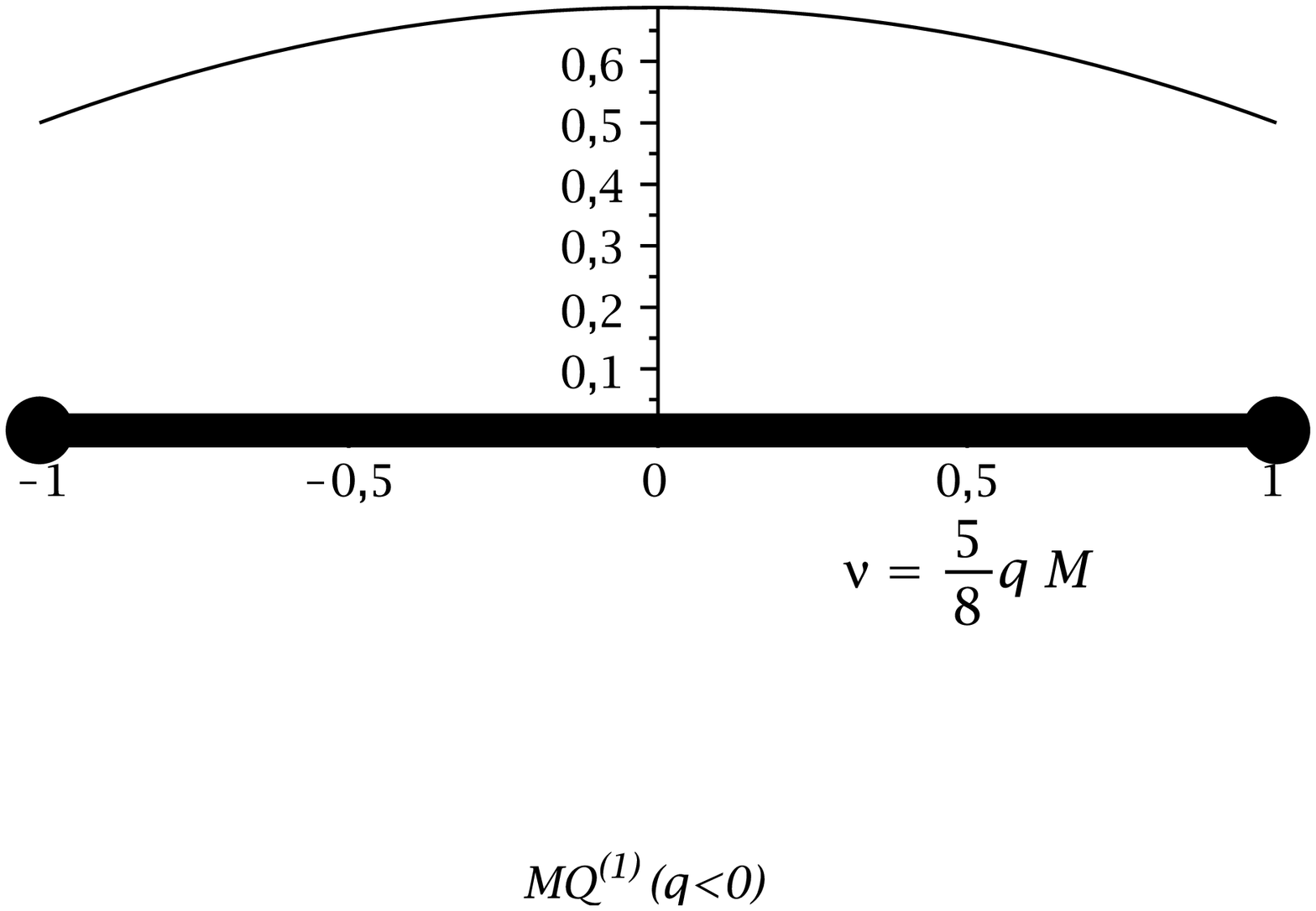,height=2.1in}
&
\epsfig{figure=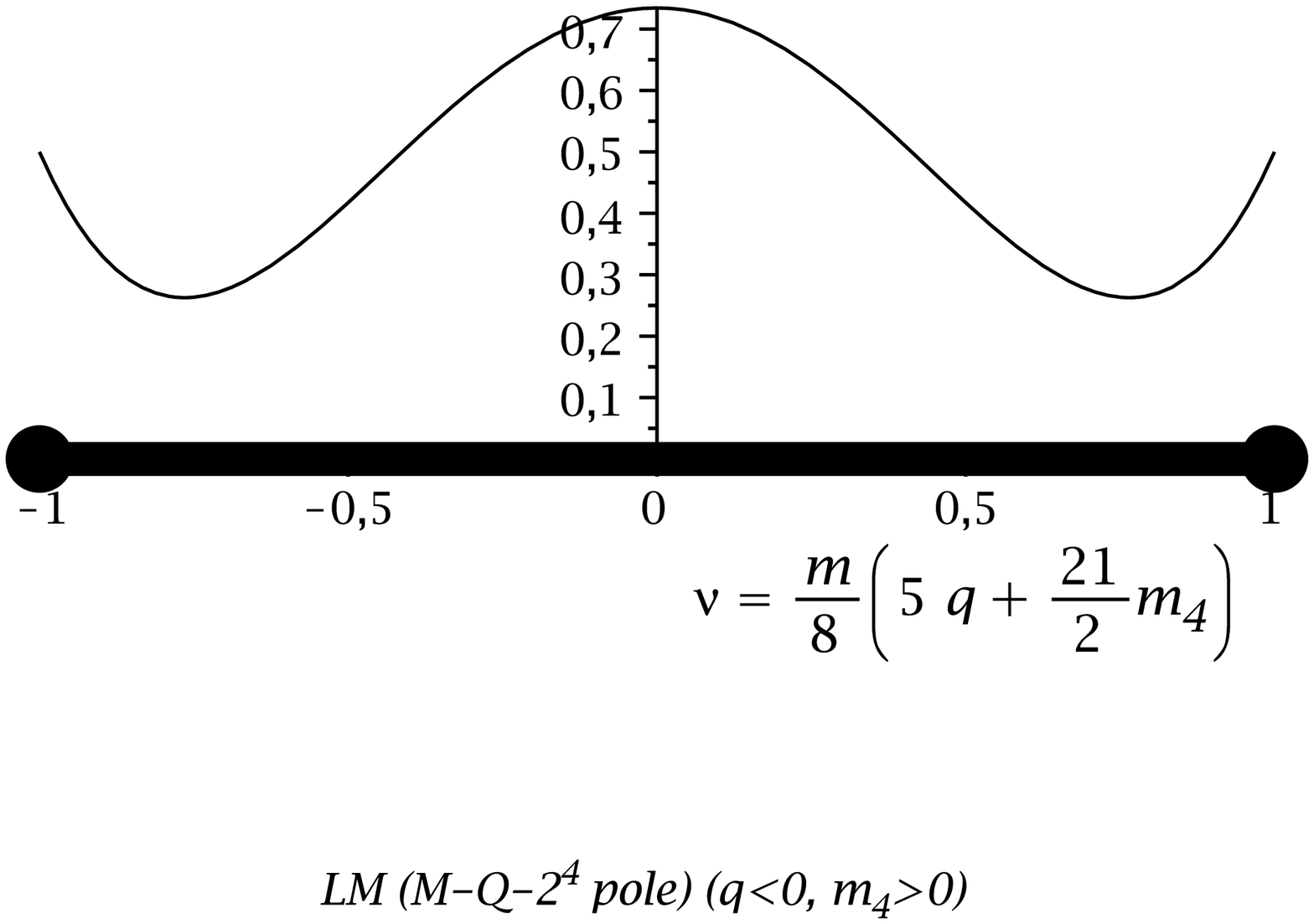,height=2.1in}\nonumber
\end{array}
\nonumber
$$
\caption{\it The characterization of different Weyl solutions by means of either
a bar or a dumbbell. The linear density is represented as well as the mass of
the dumbbell balls for those solutions.}
\end{figure}

\section{Conclusions}

This work is devoted to finding a physical object\footnote{The object called a
dumbbell and its density have been defined from
a singular mass distribution along the Z axis, but the rod of the dumbbell  can
also be considered to be a cylinder whose width $\epsilon$ is sufficiently small relative to its length $2L$ that
 the ratio  $\epsilon/L$ is negligible.} whose Newtonian
gravitational
potential equals the metric function of some relativistic axially
symmetric solutions of the static vacuum equations. Since the static and
axisymmetric solutions
are defined by means of only one metric function ($\Psi$), our claim is that the
object and its physical
features (such as its density) can be used to
characterize the Weyl solution by supplying it with an interpretation that
generalizes the case of spherical symmetry.

This identification of the solution is, of course, dependent on the system of
coordinates but  the results obtained here acquire even more interest
because the Schwarzschild
solution
 is known to be interpreted, in Weyl coordinates, as a finite bar of constant
density. We focus our search  on relativistic solutions with a well known
physical
meaning in the sense of being slight
deviations from  spherical symmetry. To this end,
  a new static and axisymmetric solution of the Einstein vacuum
equations is presented: the Linearized Multipole (LM) solutions, which represent
 approximations, linear in the  RMM, to the
general
 Pure Multipole solutions, which are defined as those having
a finite number of RMM. The particular case of  two RMM alone is the M-Q
solution
\cite{mq}.
In NG there is no interaction between the multipole moments, and multipole
coupling makes no
 contribution to the gravitational potential. The LM solution is constructed from the contributions of a finite
number of RMM without any coupling between them.
Therefore, the solution can be used to describe very slight deviations from
the case of
spherical symmetry, due  not only to the quadruple moment (the M-Q$^{(1)}$
solution is recovered in that case) but also to any other higher RMMs, if
they are very small quantities.

 If we
 consider the LM solution with only the first RMM (the monopole), we recover the
Schwarzschild solution
 and consequently the density of the dumbbell bar is constant and its balls
vanish. As increasingly  higher
RMM are considered, then the density of the bar changes in accordance with those
RMM and the masses of the balls acquire a non-zero value.  We have proved
that the object called a {\it dumbbell} is able to describe the LM solutions.
For instance, the flattened or elongated form of the source of a static and
axisymmetric space-time is related to the behaviour of the dumbbell density. We
have shown that in this
representation the horizon ($x=1$) of the LM solution  is reduced  to the region along the $Z$ axis where the rod of the dumbbell is
located  and the behaviour of the $g_{00}$ component of the metric on that
surface can be described  by means of the density of the dumbbell.
 The dumbbell representation allows us to identify the deformations from
spherical symmetry
due to different RMM in GR by means of  the physical characteristics of a
Newtonian
object.

The definition of the dumbbell and its density allows us to construct an
alternative representation of some Weyl solutions: a simple polynomial with
 $g$ independent terms describes the relativistic solution with $g+1$ multipole
moments (starting from the Monopole). The representation of this kind of
 solution with a finite number of RMM is a clear achievement, and represents an
improvement with respect  to the Weyl or ERQ representations.
Recall that not all Weyl solutions can be identified with the
potential of a dumbbell, so the search for an object  able to
describe other
solutions may be a matter of consideration for future work. Moreover, we might
consider extended Newtonian objects
 rather than singular sources to represent the  gravitational potential in
particular cases.
In fact, we should point out that it is possible to obtain general results about
 the existence of an even
density with prescribed moments like those of equation (\ref{mNG}):
\begin{equation}
 M_{2n}=\int_{-1}^1z^{2n}\mu(z)
dz=\int_0^1w^n\left[\frac{\mu(\sqrt{w})}{\sqrt{w}}\right] dw \ ,
\label{hausm}
\end{equation}
 since Hausdorff \cite{hauss} proved a set of necessary and sufficient
conditions
for the existence of a positive
function $f$ with prescribed  half-range {\it moments} $b_n$ in the sense of
equation (\ref{hausm})
$b_n=\int_0^1w^n f(w)dw$ that involve   conditions  on the  moments (we are referring to the classical problem in analysis
called the Hausdorff Moment Problem.)

As a line of enquiry, future work might be devoted to studying what other properties or
features  of the relativistic solutions can be described in terms of the density
of a dumbbell bar. In addition,
 methods for integrating the equations for the second metric function $\gamma$ in the LM solutions deserve detailed study \cite{teixeira},
as does the determination
of circular
orbits for a test particle around a source with this kind of vacuum metric.

\section{Appendix}

\subsection{Newtonian multipole moments and gravitational potential associated
with Letelier's definition of linear density}

Let us calculate the Newtonian
gravitational potential, as well as the  multipole moments
of a bar with density $\lambda(z)$ (\ref{densilete}) and length $2L$,
according to the expressions (\ref{phiNG}) and
(\ref{mNG}). On the one hand, the gravitational potential is as
follows\footnote{Let us note
that
${\displaystyle \frac{1}{\sqrt{\rho^2+(z-x)^2}}=\frac 1r
\sum_{n=0}^{\infty}\left(\frac
xr\right)^n P_n(z/r)}$.}:
\begin{equation}
 \Phi=\int_{-L}^{L} dz^{\prime} \frac{\lambda(z^{\prime})}{r}
\sum_{n=0}^{\infty}\left(\frac{z^{\prime}}{r}\right)^n
P_n(\omega)=\int_{-1}^1 \lambda(L t)
\sum_{n=0}^{\infty}t^n s^{n+1} P_n(\omega) dt \ ,
\end{equation}
where $r\equiv\sqrt{\rho^2+z^2}$, $s\equiv L/r$ and the
change of
variable $z^{\prime}=L t$ has been performed. If we introduce
the density (\ref{densilete}) into this
expression,
then the gravitational potential turns out to be:
\begin{equation}
 \Phi=\frac 12\sum_{n,k=0}^{\infty} P_n(\omega)\int_{-1}^1 s^{n+1} t^n q_kP_k(t)
dt=
\sum_{n,k=0,k\leq n}^{\infty} q_k s^{n+1}P_n(\omega)\frac{C_{n,k}}{2k+1} \ ,
\label{phibar}
\end{equation}
where $C_{n,k}$ are the coefficients appearing in the series expansion  of the
$n^{th}$
power of any variable as a linear combination
of Legendre polynomials in that variable: ${\displaystyle
\xi^n=\sum_{k=0}^{\infty} C_{n,k}P_k(\xi)}$. In \cite{mq},\cite{tesis} we
explicitly obtained
 the relation
between the sets of coefficients associated with both the ERQ (\ref{psiprola}) and
the
Weyl (\ref{psi}) representations of the axially
symmetric static vacuum solutions:
\begin{equation}
 q_{2k}=(4k+1)\sum_{j=0}^k \frac{L_{2k,2j}}{-M^{2j+1}} a_{2j} \quad , \quad
q_{2k+1}=0 \ ,
\label{relaqa}
\end{equation}
where $L_{2k,2j}$ is the coefficient multiplying the $2j$ power  of the variable
$\xi$ in the Legendre polynomial $P_{2k}(\xi)$.
Hence,  the Newtonian potential (\ref{phibar}) transforms into the following
expression after taking (\ref{relaqa}) into
account:
\begin{eqnarray}
 \Phi&=&-\sum_{n=0}^{\infty} \sum_{k=0}^n s^{2n+1}
P_{2n}(\omega)C_{2n,2k}\sum_{j=0}^k L_{2k,2j}\frac{a_{2j}}{M^{2j+1}}=
\nonumber\\
&-&\sum_{n=0}^{\infty}  s^{2n+1} P_{2n}(\omega)\sum_{j=0}^n
\frac{a_{2j}}{M^{2j+1}} \sum_{k=j}^n C_{2n,2k} L_{2k,2j} \ ,
\label{phibar2}
\end{eqnarray}
and so we can finally conclude that\footnote{Let us note that we take
$L=M$ and we have used the relation (see \cite{tesis} for details):
 ${\displaystyle \sum_{k=j}^n C_{2n,2k} L_{2k,2j} = \delta_{jn}}$.}
the Newtonian gravitational potential $\Phi$ is equal to the metric function
$\Psi$ of the Weyl family of solutions (\ref{psi}):
\begin{equation}
 \Phi=\Psi=-\sum_{n=0}^{\infty} r^{-(2n+1)} P_{2n}(\omega)a_{2n} \ .
\end{equation}

On the other hand, in terms of the multipole moments of the bar, we can
write, according to (\ref{mNG}), the following expression:
\begin{equation}
M_{2n}^{NG}=(\pi\epsilon^2)\int_{-1}^1  L^{2n+1} t^{2n}
\lambda(L t) dt \ ,
\label{mNGbar1}
\end{equation}
and the introduction of the density $\lambda(z)$ (\ref{densilete}) into the
integral leads to
\begin{equation}
 M_{2n}^{NG}=\frac 12 (\pi\epsilon^2) L^{2n+1} \sum_{k=0}^{\infty} q_k
\int_{-1}^1  t^{2n} P_k(t) dt \ ,
\label{mNGbar2}
\end{equation}
and we can conclude with the following identity\footnote{We have used the following relation:
${\displaystyle \int_{-1}^1  t^{2n} P_k(t) dt= \frac{2C_{2n,2j}}{4j+1}}$ iff
$k=2j$ (even) with $C_{2n,2j}=0$ for $j>n$.
In the case of an odd $k$ ($k=2j+1$) the integral is zero.}:
\begin{equation}
 M_{2n}^{NG}=(\pi\epsilon^2) L^{2n+1} \sum_{j=0}^{n} q_{2j}
\frac{C_{2n,2j}}{4j+1}=-a_{2n} \ .
\label{mNGbar3}
\end{equation}
In other words, we recover the inverse relation of (\ref{relaqa}) (see \cite{mq1} for details) existing
between
the  sets of coefficients associated with both the ERQ (\ref{psiprola}) and the
Weyl
(\ref{psi}) representations of the axially
symmetric static vacuum solutions, and hence we can say that the
Newtonian multipole moments of the bar with
density given by $\lambda(z)$ (\ref{densilete}) are simply the coefficients
$a_{2n}$ of the Weyl family of solutions.

\section{Acknowledgments}
This  work  was partially supported by the Spanish  Ministerio de Ciencia e
Innovaci\'on under Research Project No. FIS2012-30926, and the Consejer\'\i a
de
Educaci\'on of the Junta de Castilla y Le\'on under the Research Project Grupo
de Excelencia GR234.

\end{document}